\documentstyle[preprint,aps,epsfig]{revtex}
\begin{document}
\draft
\preprint{KAIST-CHEP-96/01}
\title{ Heisenberg-picture approach to the exact quantum motion
	of a time-dependent forced harmonic oscillator
	}
\author{   
Hyeong-Chan Kim\thanks{Electronic address: leo@chep5.kaist.ac.kr}, 
Min-Ho Lee\thanks{Electronic address: mhlee@chep6.kaist.ac.kr}, 
Jeong-Young Ji\thanks{Present Address: Department of Physics
Education, Seoul National University, Seoul, 151-742, Korea.
Electronic address: jyji@phyb.snu.ac.kr},
and Jae Kwan Kim  
}
\vspace{.5in}
\address{           
Department of Physics, Korea Advanced Institute of Science and Technology, 
Taejon 305-701, KOREA
}
\date{\today}
\maketitle

\begin{abstract}
In the Heisenberg picture, the generalized invariant and
exact quantum motions are found for a time-dependent forced
harmonic oscillator.
We find the eigenstate and the coherent state of the invariant
and show that the dispersions of these quantum states
do not depend on the external force.
Our formalism is applied to several interesting cases.
\end{abstract}

\narrowtext
\newpage

\section{Introduction}

The quantization of time-dependent harmonic oscillator is very important when
we treat the coherent states, the squeezed states,
and many other branches of physics. For example, the propagator
of the quantum mechanical system with friction can be treated as 
a harmonic oscillator with time dependent mass \cite{jann}. 
One of the most powerful method to find a quantum state of 
the oscillator is 
the generalized invariant method introduced by Lewis and 
Riesenfeld (LR)~\cite{lewis}.
Recently, some of the authors obtained the most general form of 
the invariant and the general quantum states
for the oscillator using the Heisenberg picture when the mass
and the frequency are explicitly time dependent~\cite{sangpyo,jyji1}.

The time-dependent harmonic oscillator with a perturbative force was
considered by Khandekar and Lawande \cite{kha} who found that 
the wave function and propagator depend only on the solution of a
classical damped oscillator through a single function.  
The exact evolution operator for the harmonic oscillator
subject to arbitrary force was obtained by Lo~\cite{lo}.
He obtained the quantum states by solving the differential equation
satisfied by the evolution operator. He also investigated the time evolution 
of a charged oscillator with a time dependent mass and frequency 
in a time-dependent field.

We consider the forced harmonic oscillator,
where the external force depends on time.
Even though Lo found the solution in the general situation,
the solutions are difficult to analyze the physical properties of
the system. While, our formulations has an advantage in that point,
since the classical solutions suffice to
understand the corresponding quantum system.
Our choice of quantum state is different from 
those of Ref.~\cite{kha,lo}. 
They choose the Fock space in terms of 
the eigenstates of the Hamiltonian with the external force term suppressed. 
However, its ground state cannot be the lowest-energy state 
since the external force lower the minimum of the potential,
hence the zero-point energy. 
Further, the dispersions of $p$ and $q$ for those states
are dependent on the external force. 

On the other hand, we construct the Fock space in terms of  
the eigenstates of the generalized invariant for {\it the forced 
oscillator.}
These number-states can be set equal to the ones of the Hamiltonian
by choosing the parameter properly. It should be emphasized that
the dispersions of $p$ and $q$ for these states are independent of the
external forces.

In Sec.~II  we present the general formulation for finding the quantum
solutions of the time-dependent forced harmonic oscillator 
using the LR invariant method.
In Sec. III we find
the quantum evolution of the Heisenberg operator, $q(t)$ and $p(t)$,
and the time-evolution operator.
In Sec. IV we take some examples for the applications and
Sec. V is devoted to summary.

\section{Forced harmonic oscillator  -   General formulation}

The Hamiltonian for a time-dependent harmonic oscillator under a force $F(t)$
is 
\begin{eqnarray} 
H_T(t) = H(t)+ H_F(t) = \frac{p^2(t)}{2M(t)} + 
			 \frac{M(t) \omega^2(t)}{2}q^2(t) - M(t)F(t) q(t).
\end{eqnarray}
In the previous paper \cite{jyji1} we found the invariant $I(t)$ 
which satisfies
\begin{equation}
\frac{\partial}{\partial t} I(t) -i [I(t), H(t)] =0.
\end{equation}
The invariant $I(t)$ is 
\begin{eqnarray}  \label{I:g}
I(t) = g_-(t) \frac{p^2(t)}{2} + g_0(t)\frac{p(t)q(t)+ q(t)p(t)}{2}
	+ g_+(t) \frac{q^2(t)}{2},
\end{eqnarray}
and $g_k(t)$ are given  in Ref.~\cite{jyji1}
using two independent solutions of
\begin{eqnarray}
\frac{d}{dt} \left[ M(t) \frac{d}{dt} f_{1,2}(t)\right]
		+ M(t) \omega^2(t) f_{1,2}(t) =0.
\end{eqnarray}
Introducing the creation and annihilation operators:
\begin{eqnarray} \label{b(t)}
b(t) &=& \left( \sqrt{\frac{\omega_I}{2g_-(t)} }+ i \sqrt{\frac{1}{2\omega_I 
		g_-(t)}} g_0(t)
	 \right) q(t) + i\sqrt{\frac{g_-(t)}{2 \omega_I}} p(t), \\
b^{\dagger}(t) &=& 
	 \left( \sqrt{\frac{\omega_I}{2g_-(t)} }- i \sqrt{\frac{1}{2\omega_I 
		g_-(t)}} g_0(t)
	 \right) q(t) - i\sqrt{\frac{g_-(t)}{2 \omega_I}} p(t),
\end{eqnarray}
we can write the invariant as 
\begin{eqnarray} \label{I:b}
I(t) = \omega_I \left( b^{\dagger}(t) b(t) +\frac{1}{2}\right),
\end{eqnarray}
where $\omega_I^2  =  g_+(t) g_-(t) - g_0^2(t)$ is a constant of motion.

In the presence of the external force, $I(t)$ is no longer the invariant. 
therefore, we search for the invariant of the form
\begin{eqnarray}
I_T(t) = \omega_I 
      \left( B^\dagger (t)B (t) + \frac{1}{2} \right),
\label{I:B}
\end{eqnarray}
where 
\begin{eqnarray}
B(t)=b(t) + \beta(t).
\label{B:b}
\end{eqnarray}
Here $\beta(t)$ is a c-function of time and
$[B(t), B^\dagger(t)]= 1$.

The Heisenberg equation of motion for $b(t)$ leads to 
\begin{eqnarray}
\frac{d}{d t} b(t) &=& \frac{\partial}{\partial t} b(t) 
			- i [b(t), H(t) + H_F(t)] \\
 	&=& -i \frac{\omega_I}{M(t) g_-(t)} b(t) + i M(t) F(t) 
		\sqrt{ \frac{g_-(t)}{2 \omega_I }}. \nonumber
\end{eqnarray}
Then the Heisenberg equation for $B(t)$ can be written as
\begin{eqnarray} \label{a2}
 \frac{d}{dt} B(t) =
     -i \frac{\omega_I}{M(t) g_-(t)} B(t), 
\end{eqnarray}
where $\beta(t)$ satisfies the differential equation    
\begin{eqnarray} \label{beta:0}
\frac{d}{dt} \beta(t) +i \frac{\omega_I}{M(t) g_-(t)} \beta(t)
	= -i M(t) F(t) \sqrt{\frac{g_-(t)}{2 \omega_I}}.
\end{eqnarray}
Further its solution is easily found
\begin{eqnarray} 
\beta(t) = e^{-i \Theta(t)}\beta_0-i e^{-i\Theta(t)} 
	 \int^{t}_{t_0} dt 
	      \sqrt{\frac{g_-(t)}{2 \omega_I}} M(t) F(t) 
	      e^{i\Theta(t)} ,
	      \label{h:F}
\end{eqnarray}
where $\beta_0$ is an arbitrary parameter which will be fixed 
by the extra condition.
For the later convenience, we define
\begin{eqnarray}
{\cal F}(t) = e^{-i \Theta(t)}\beta_0- \beta(t). 
\end{eqnarray}
Now, the solutions of (\ref{a2}) are
\begin{eqnarray}\label{b1}
B^{\dagger}(t) = e^{i \Theta(t)} B^{\dagger}(t_0), ~~~~~ 
B(t) = e^{-i \Theta(t)} B(t_0),
\end{eqnarray}
where 
\begin{eqnarray}
\Theta(t) = \int^{t}_{t_0}dt \frac{\omega_I}{M(t) g_-(t)} .
\label{Theta:t}
\end{eqnarray}
Thus we have found the generalized invariant even in the presence 
of the external force. As seen from (\ref{b(t)}) and (\ref{B:b}), 
the creation and annihilation operators of the invariant
are constructed from shifting those of the corresponding Hamiltonian 
with no external force. 

From Eq. (\ref{b1}), we can easily notice that (\ref{I:B}) is 
indeed the invariant for $H_T(t)$
and it can be rewritten as 
\begin{eqnarray}
I_T(t) &=& \frac{1}{2} \omega_I [ B(t) B^{\dagger}(t) + B^{\dagger}(t) B(t)] \\
     &=& I(t)+ \omega_I Re(\beta) \sqrt{\frac{2 \omega_I}{g_-(t)}} q(t)
	 + \omega_I Im(\beta ) \sqrt{\frac{2g_-(t)}{ \omega_I}} \left( p(t) + 
	 \frac{g_0(t)}{g_-(t)} q(t) \right) + \omega_I|\beta(t)|^2. \nonumber
\end{eqnarray}
By setting the parameter constant $\beta_0$ in Eq.~(\ref{h:F}) as 
\begin{eqnarray}
\beta_0 = -\frac{1}{2} \frac{M(t_0)}{\omega_I}\sqrt{\frac{2 g_-(t_0)}{\omega_I}}
	F(t_0) 
\end{eqnarray}
and choosing $I(t)$ according to Eq. (3.4) of Ref.~\cite{jyji3},
the invariant becomes the instantaneous Hamiltonian at $t_0$ with the
additional constant, i.e.
\begin{eqnarray}\label{I:Tbeta}
I_T(t_0) = H_T(t_0)+ \beta_0^2. 
\end{eqnarray}

Let us define the creation and annihilation operator of the
Hamiltonian with no external force by
\begin{eqnarray}\label{H:a}
H(t)=\omega(t) \left( a^\dagger (t) a(t) + \frac{1}{2}
\right).
\end{eqnarray}
Then the vacuum state $\left| 0 , t \right>_a$ is defined by 
\begin{eqnarray}
a(t)\left|0\right>_a= 0, 
\label{0:a}
\end{eqnarray}
and the number-states are constructed  as
\begin{eqnarray}
\left| n, t \right> = \frac{a^{\dagger n} (t)}{\sqrt{n!}}
\left| 0, t \right>. 
\label{n:a}
\end{eqnarray}
These states were studied by Lo~\cite{lo}.

Here, we construct the Fock space as eigenstates of $I_T(t)$.
The vacuum state $\left|0\right>_B$  for the harmonic oscillator
is defined by
\begin{eqnarray}
   B(t) \left| 0 \right>_B =0,
   \label{0:B}
\end{eqnarray}
and the number-states by
\begin{eqnarray}
\label{n:B}
\left|n, t\right>_B = \frac{{B^{\dagger}}^n(t)}{\sqrt{n !}}
				\left|0, t\right>_B.
\end{eqnarray}
It can be written as
\begin{eqnarray}
\left|n, t\right>_B = e^{i (n + 1/2 )\Theta(t)} \left|n, t_0\right>_B,
\end{eqnarray}
with the phase of $\left|0\right>_B$ fixed properly.

These are different 
from (\ref{0:a}) and they are correlated through the
following Bogoliubov transformation:
\begin{eqnarray}
B(t) = v_1 (t) a(t) + v_2 (t) a^{\dagger}(t) + \beta(t),
\end{eqnarray}
where 
\begin{eqnarray}
v_1(t) &=& 
\frac{1}{2} \left[
\sqrt{\frac{M(t)g_- (t) \omega(t)}{\omega_I}}
+\sqrt{\frac{\omega_I}{M(t) g_- (t) \omega(t)}}
\left(1+i \frac{g_0 (t)}{\omega_I}\right)\right],
\nonumber \\
v_2(t) &=& 
\frac{1}{2} \left[
-\sqrt{\frac{M(t)g_- (t) \omega(t)}{\omega_I}}
+\sqrt{\frac{\omega_I}{M(t) g_- (t) \omega(t)}}
\left(1+i \frac{g_0 (t)}{\omega_I}\right)\right],
\end{eqnarray}
with $ |v_1(t)|^2 - |v_2(t)|^2 = 1$.
This transformation can be written as the unitary one  
\begin{eqnarray}
B(t) = S^{\dagger}(t) D^{\dagger}(t) a(t) D(t) S(t).
\end{eqnarray}
Here the squeezing operator is given by 
\begin{eqnarray}
S(t) = \exp(i \theta_{1} a^\dagger a)\exp\left(\frac{1}{2}
	e^{i(\theta_{2}- \theta_{1})} \cosh^{-1}|v_1| a^{\dagger 2}
	- H.C. \right),
\end{eqnarray}
where $v_1=|v_1| e^{i \theta_{1}}$, $v_2=|v_2| e^{i \theta_{2}}$,
and the displacement operator is
\begin{eqnarray}
D(t) = e^{\beta(t) a^{\dagger}(t)-\beta^*(t) a(t)}.
\end{eqnarray}
Thus the ground state of (\ref{n:B}) are the displaced squeezed states of
(\ref{n:a}), and vice versa:
\begin{eqnarray}
\left|0, t\right>_B = S^{\dagger}(t) D^{\dagger}(t) \left|0,t\right>_a.
\label{0B:0a}
\end{eqnarray}
Furthermore, it should be noted that the vacuum state of (\ref{n:B}) is
the coherent state of the unforced oscillator:
\begin{eqnarray}\label{0B:0b}
\left| 0 \right>_B   
=   e^{-|\beta(t)|^2/2} 
   \sum_{n=0}^{\infty} \frac{|\beta(t)|^2}{n!} b^{\dagger n}
\left| 0 \right>_b,
\end{eqnarray}
with
\begin{eqnarray}
 b \left| 0 \right>_{b} = 0.
\end{eqnarray}

Now let us calculate the vacuum expectation value of $H_T(t)$ for 
$\left|0\right>_a$ and $\left|0,t\right>_B$.
The Hamiltonian can be rewritten by  $B(t)$, $ B^{\dagger}(t)$,
and $\beta(t)$ as:
\begin{eqnarray}
H_T(t) = H_Q(t) -d(t) B^{\dagger}(t) - d^{\dagger}(t) B(t)
		+ H_d(t),
\end{eqnarray}
where  $H_Q(t)$ is the quadratic term in $B(t)$ and $B^{\dagger}(t)$,
and $H_d(t)$ is quadratic in $F(t)$.
\begin{eqnarray}
H_Q(t) &=& h_+(t) \frac{{B^\dagger}^2(t)}{2}+h_0(t) \frac{B(t)B^\dagger(t)+
	   B^\dagger(t)B(t)}{4} + h_-(t)\frac{{B(t)}^2}{2},\\
H_d(t) &=& h_+(t) \frac{\beta^2(t)}{2} + h_0(t) \frac{\beta(t)\beta^\dagger(t)+
           \beta^\dagger(t) \beta(t)}{4} 
	   + h_-(t) \frac{{\beta^\dagger}^2(t)}{2}  \\
       &-&M(t) \sqrt{\frac{g_-(t)}{2 \omega_I}} F(t)\left[\beta(t) 
	  + \beta^\dagger(t)\right], \nonumber
\end{eqnarray}
where $h_i$ is defined at the previous paper \cite{jyji1} and 
the displacement $d(t)$  
\begin{eqnarray}
d(t) &=& h_0(t) \beta(t) + h_-(t) \beta(t) 
	+ \sqrt{\frac{g_-(t)}{2\omega_I}} M(t)F(t),
\end{eqnarray}
and $H_d(t)$ are proportional to the identity operator.
From (\ref{H:a}) and ( \ref{0:a}) the vacuum expectation value
of $H_T(t)$ for $\left|0\right>_a$ is
\begin{eqnarray}
_a \left< 0 \right| H_T(t) \left| 0\right>_a = \frac{\omega(t)}{2}.
\end{eqnarray}
On the contrarily, if we take the instantaneous invariant states
from  (\ref{I:Tbeta})  and (\ref{0:B}),
the vacuum expectation value for $\left|0 \right>_B$
becomes
\begin{eqnarray}
_B \left<0 ,t_0\right| H_T(t_0) \left| 0, t_0 \right>_B=\frac{\omega(t_0)}{2}
          -  \beta_0^2.
\end{eqnarray}
The state (\ref{0:a}) cannot be the lowest energy state as expected;
the minimum of the quadratic potential is lowered by the external force.

\section{Quantum Evolutions}

Now let us find the quantum evolutions of $p$ and $q$ which are
the only things we should to know in order to study the quantum
mechanical system in the Heisenberg picture. 
Equating the Hermitian and anti-Hermitian parts of both sides of (\ref{b1})
and using (\ref{b(t)}) and (\ref{B:b}),
the time evolution of $q(t)$ and $p(t)$ is given by
\begin{eqnarray}
q(t) &=&  q(t_0) \sqrt{\frac{g_-(t)}{g_-(t_0)} } \left[ \cos \Theta(t)+
          \frac{g_0(t_0)}{\omega_I} \sin \Theta(t)\right] 
	+ p(t_0) \frac{ \sqrt{g_-(t)g_-(t_0)}}{\omega_I} \sin\Theta(t) \nonumber\\
     &+& \sqrt{\frac{g_-(t)}{2\omega_I}} [{\cal F}(t)
	 +{\cal F}^{\dagger}(t)],  \label{q:t}\\
p(t) &=&  q(t_0) \frac{1}{ \sqrt{g_-(t)g_-(t_0)}} 
	\left[ \{g_0(t_0)-g_0(t)\} \cos \Theta(t) -\left\{ \omega_I +
	\frac{g_0(t_0)g_0(t)}{\omega_I} \right\} \sin \Theta(t) \right] \nonumber\\
     &+& p(t_0)\sqrt{\frac{g_-(t_0)}{g_-(t)}} \left\{ 
	  \cos \Theta(t) - \frac{g_0(t)}{\omega_I} \sin \Theta(t) \right\}\\
     &-&i\sqrt{\frac{\omega_I}{2g_-(t)}}\left[ \left\{
         1-i\frac{g_0(t)}{\omega_I}  \right\} {\cal F}(t)
	- \left\{ 1+ i\frac{g_0(t)}{\omega_I} \right\} {\cal F}^\dagger(t)
			   \right].  \label{p:t}  \nonumber
\end{eqnarray}
By direct calculation, 
one can easily check that (\ref{q:t}) and (\ref{p:t}) satisfies the equation of
motion 
\begin{eqnarray}
\frac{d q(t)}{d t} = -i \left[q(t), H(t)\right] ,     \\
\frac{d p(t)}{dt}  = -i \left[p(t), H(t)\right]. \nonumber
\end{eqnarray}

The dispersions of $p$ and $q$  for a number state (\ref{n:B}) are given by
\begin{eqnarray}
_B \left<n \right|[\Delta q(t)]^2\left|n \right>_B
    &=&(2n +1)\frac{g_-(t)}{2 \omega_I},  \label{Deltaq:B}\\
_B \left<n \right|[\Delta p(t)]^2\left| n \right>_B 
    &=& (2n+1)\frac{\omega_I}{2g_-(t)} 
			\left[1+ \frac{g_0^2(t)}{\omega_I^2}\right].
			\label{Deltap:B}
\end{eqnarray}
If we construct the coherent state of the forced oscillator as
\begin{eqnarray}\label{coher:B}
\left|\alpha \right> =  e^{-|\alpha|^2/2}
	\sum_{n=0}^{\infty} \frac{|\alpha|^2}{n!}
	B^{\dagger n} \left| 0 \right>_B,
\end{eqnarray}
with $\alpha = |\alpha| e^{-i\delta}$,
the dispersions in $q$ and $p$ for this state is given by (\ref{Deltaq:B})
and (\ref{Deltap:B}) with $n=0$.
It should be noted that the dispersions (\ref{Deltaq:B}) 
and (\ref{Deltap:B}) do not depend on $F(t)$ and
they are the same as the ones of the eigenstate of the invariant (\ref{I:b})
with no external force. Thus the shape of the wavepacket for the eigenstate
of the invariant does not alter by the external force. The physical meaning
of this result is clear and non-frightening: the force $F(t)$ 
acts uniformly on the wavepacket so that the wavepacket is shifted
with no distortion. In other words, the eigenstate of the invariant
is the quantum state whose wavepacket does not alter its shape
by the external force.
However, a general state which is constructed
by a linear combination of the eigenstates does not possess this
property because of the time-dependent phase. This phase makes 
interferences between the eigenstates with different quantum numbers.
However, the coherent state (\ref{coher:B}), although it is made 
by the linear combination of the eigenstates, has the same property as
the eigenstate. This is a remarkable property of a coherent state.

Further the expectation value $\left<q(t)\right>$ and 
$\left<p(t)\right>$ becomes
\begin{eqnarray}
\left< \alpha \right| q(t) \left|\alpha \right> &=&  
	\sqrt{\frac{g_-(t)}{2 \omega_I}} \left\{ 2 |\alpha|
	\cos[\Theta(t)+ \delta] - {\cal F}(t) -{\cal F}^\dagger(t) \right\}\\ 
\left< \alpha \right| p(t) \left|\alpha \right> &=&
	\sqrt{\frac{\omega_I}{2 g_-(t)}} \left\{
	- 2|\alpha| \sin[\Theta(t)+\delta]-
	  \left(i+\frac{g_0(t)}{\omega_I }\right){\cal F}(t)
	  -\left(i-\frac{g_0(t)}{\omega_I}\right) {\cal F}^\dagger(t) 
	\right\}.
\end{eqnarray}
Thus the wavepacket of the coherent state moves backward and forward 
like a classical particle. 

Finally we want to find the time evolution operator $U_T(t, t_0)$.
In the Heisenberg picture the position and momentum operators transform
according to
$q(t) = U_T^{\dagger} (t, t_0) q(t_0) U_T(t,t_0), $ and $
p(t) = U_T^{\dagger} (t, t_0) p(t_0) U_T(t,t_0).$
By expressing these in terms of the creation and annihilation operators
at $t_0$ and using the Eqs. (\ref{b(t)}) and (\ref{b1}), we obtain
\begin{eqnarray}\label{aeq}
U_T^{\dagger} (t, t_0) B(t_0)  U_T(t,t_0)
 &=& - u_2^*(t) B^{\dagger}(t_0) e^{i\Theta(t)}
	 +u_1(t) B(t_0) e^{-i\Theta(t)} + d(t), \\
     U_T^{\dagger} (t, t_0) B^{\dagger}(t_0)  U_T(t,t_0)
 &=&  u_1^*(t) B^{\dagger}(t_0) e^{-i\Theta(t)}  \label{Bdagger:b}
	  -u_2(t) B(t_0) e^{i\Theta(t)} + d^*(t),
\end{eqnarray}
where $u_1(t)$ and $u_2(t)$ are given \cite{jyji1} and $d(t)$ is
\begin{eqnarray}
d(t) &=& \frac{1}{2} \left[ \left( 1+ i \frac{g_0(t_0)}{2 \omega_I}
  \right) \sqrt{\frac{g_-(t)}{g_-(t_0)}} \left\{\beta(t)
  + \beta^{\dagger}(t) \right\} \right.\\
 &+& \left. \sqrt{\frac{g_-(0)}{g_-(t)}} \left[
 \left\{\beta(t) - \beta^\dagger(t)\right\}-i \frac{g_0(t)}{\omega_I}
  \left\{\beta(t) + \beta^\dagger(t)\right\} \right]
 \right]^{t}_{t_0}. \nonumber
\end{eqnarray}
Then the time-evolution operator $U_T (t,t_0 )$ is given by
\begin{eqnarray}
U_T(t,t_0) = U_d(t,t_0) U(t,t_0),  \label{UT:t}
\end{eqnarray}
where
\begin{eqnarray}
 U_d(t,t_0)= e^{d B^{\dagger}(t_0) - d^* B(t_0)} 
\end{eqnarray}
is the displacement operator and
$U(t,t_0)$ is given by
\begin{eqnarray}
U(t,t_0) &=& \exp\left(\frac{i}{2} \phi_1 \left\{B^{\dagger}(t_0)
      B(t_0)+B(t_0)B^\dagger(t_0)\right\} \right)  \\
 & &\exp \left(
 \frac{\nu}{2} \left\{e^{-i(\phi_1-\phi_2)} {B^{\dagger}}^2(t_0) -
 e^{i(\phi_1-\phi_2)} B^2(t_0)\right\} \right) , \nonumber
\end{eqnarray}
(for the definition of $\phi_1$, $\phi_2$,
and $\nu$, see Ref.~\cite{jyji1}).
It is easily checked that (\ref{UT:t}) indeed satisfies (\ref{aeq}) 
by using the following properties of the displacement operator
\begin{eqnarray}
U_d^{\dagger}(t,t_0) B(t_0) U_d(t,t_0) &=& B(t_0) + d(t).\\
U_d^{\dagger}(t,t_0) B^{\dagger}(t_0) U_d(t,t_0)&=& B^{\dagger}(t_0) + d^*(t).
\end{eqnarray}
Thus we obtained the time-evolution operator for the forced
harmonic oscillator. This operator is in agreement with that of Lo~\cite{lo}.

\section{Some exactly solvable models}

The simplest example is the harmonic
oscillator under the constant driving force $F$. Let all
quantities are time independent
$M(t) = m$, and $\omega_0(t) =\omega$. After a little algebra,
we get the following results:
\begin{eqnarray}
q(t) - \frac{F}{\omega^2} &=& \left(q(t_0) 
		- \frac{F}{\omega^2} \right)\cos \omega(t-t_0)
		+\frac{p(t_0)}{m \omega} \sin \omega(t-t_0), \\
p(t) &=& -\frac{\omega_I}{m}\left(q(t_0)- \frac{F}{\omega^2}
		\right) \sin \omega(t-t_0)
	+p(t_0)\cos \omega(t-t_0).
\end{eqnarray}
As expected, the center of oscillation is shifted by $ F/\omega^2$
in $q$ space.

Next, let us consider the damped pulsating oscillator
with an arbitrary driving force $F(t)$, the mass
$M(t) = m_0 \exp[2(\gamma t + \mu \sin \nu t) ]$, and
the frequency $\omega^2(t) = \Omega^2 + \frac{1}{\sqrt{M(t)}} \frac{d^2
\sqrt{M(t)}}{d t^2}$. This example was also studied by Lo~\cite{lo}.
Without loss of generality, we can set the initial time $t_0 =0$.
It is easy to find the classical solution:
$f(t) = e^{i \Omega t}/\sqrt{M(t)}$ and two real independent solutions,
say $f_1(t)$ and $f_2(t)$,
can be obtained by taking its real and imaginary parts, respectively.
By setting the parameter constants in Eq.~(7) of
Ref.~\cite{jyji1} as $c_1=1=c_3,~c_2=0$, we have
\begin{eqnarray}
g_-(t) = \frac{1}{M(t)},  ~~ g_0(t) = \frac{1}{2}\frac{d}{dt} \ln M(t),
~~ g_+(t) = \frac{M(t)}{4}\left( \left[\frac{d}{dt}
	\ln M(t)\right]^2 + \Omega^2\right). \label{g:M}
\end{eqnarray}
Thus, we have fixed  the invariant (\ref{I:g}) with the frequency
$\omega_I^2 = g_+ g_- - g_0^2 = \Omega^2$. For the ground state and
the coherent state of this invariant, we have the dispersions 
in $q$ and $p$
\begin{eqnarray}
  \left< \Delta p \right>^2 &=&
  \frac{\Omega M(t)}{2} \left[ 1+ \left( \frac{\gamma +
  \mu \nu \cos \nu t}{ \Omega}
  \right)^2  \right],\\
  \left< \Delta q \right>^2 &=& \frac{1}{2 \Omega M(t)}.
\end{eqnarray}
Note that these are not affected by the
external force, as stated in the previous section.
This property makes our quantum state be distinguishable from
that of Lo -- see Eqs.~(89) and (90) in Ref.~\cite{lo}.
Furthermore, the quantum motions of $q(t)$ and $p(t)$ are obtained
by inserting (\ref{g:M}) into (\ref{q:t}) and (\ref{p:t}):
\begin{eqnarray}
q(t) &=& q(0)\sqrt{\frac{m_0}{M(t)}}
	\left[ \cos \Omega t + \frac{\gamma+ \mu \nu}{\Omega} \sin \Omega t
	\right]
	+ p(0) \frac{1}{\Omega \sqrt{m_0 M(t)}} \sin \Omega t \\
	 &+& \frac{1}{\Omega \sqrt{M(t)}}
	\int^t_0 dt' \sqrt{M(t')} F(t') \sin \Omega(t'-t),
	\label{q:0} \nonumber \\
p(t) &=& q(0) \sqrt{m_0M(t)}
	\left[ \mu \nu(1- \cos \nu t) \cos \Omega t
	- \left( \Omega + \frac{\gamma + \mu \nu}{\Omega}\right)(\gamma +
	\mu \nu \cos \nu t ) \sin \Omega t
		\right]        \nonumber \\
	 &+& p(0) \sqrt{\frac{M(t)}{m_0}} \left[\cos \Omega t -\frac{1}{\Omega}
	( \gamma + \mu \nu \cos \nu t) \sin \Omega t \right] \\
	 &-&\sqrt{M(t)} \left[ \int_0^t dt'\sqrt{M(t')} F(t')
	\cos \Omega(t'-t)\right. \nonumber\\
	 &+& \left. 2 \frac{ \gamma + \mu \nu \cos \nu t}{\Omega}
	\int_0^t dt'\sqrt{M(t')} F(t')\sin  \Omega(t'-t) \right].
	\label{p:0} \nonumber
\end{eqnarray}
Let us see the time-evolutions of the coherent states with (a)
and without (b) the external force in the 
phase space diagram (Fig. 1).
It vividly shows that the dispersion does not depends on the
external force, although their 
time-evolutions of the expectation values of $p$ and $q$ are
completely different.

\begin{figure}[bh]
\vspace{0.5cm}
\centerline{
\epsfig{figure=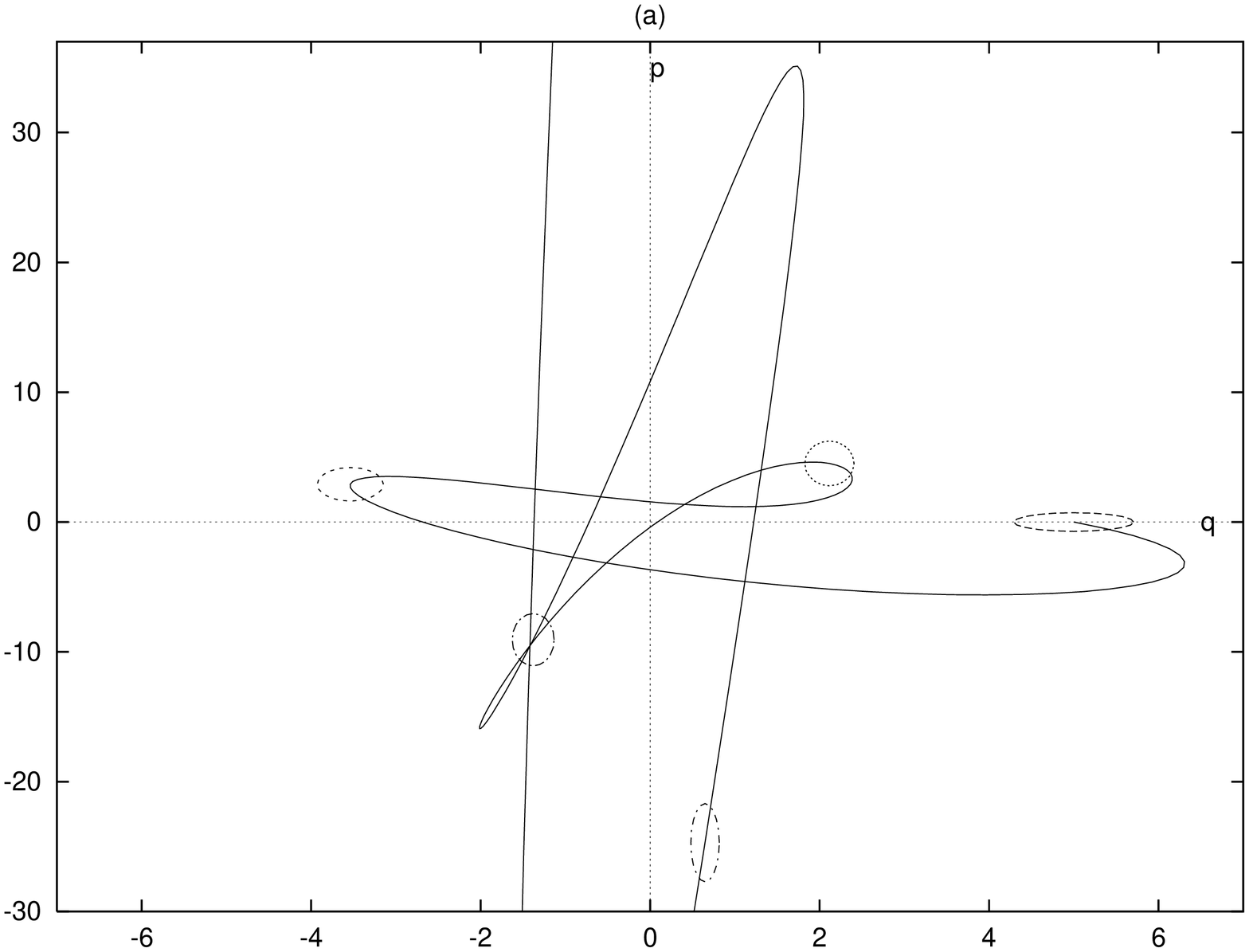,height=7cm,angle=-90}}
\vspace{0.5cm}
\centerline{
\epsfig{figure=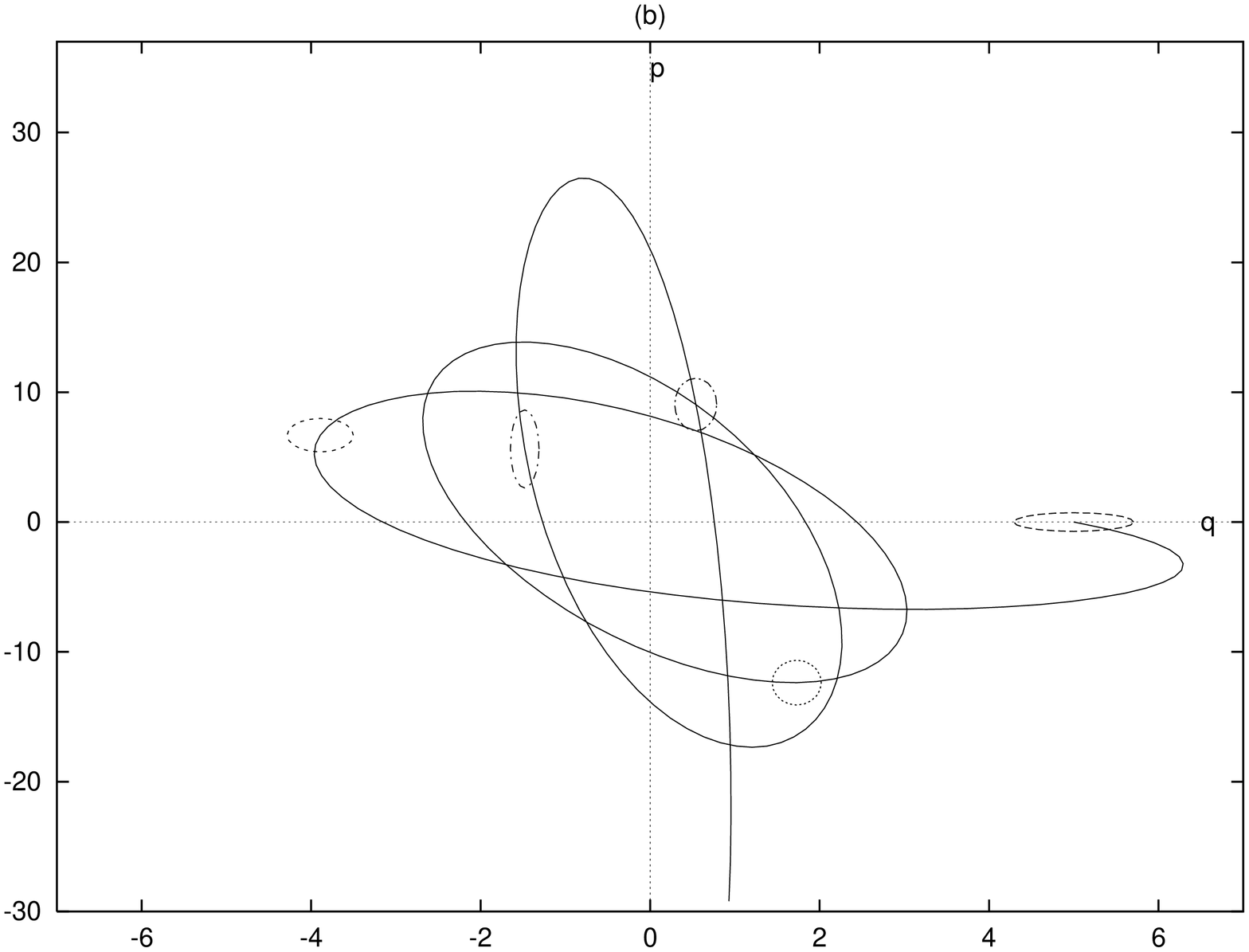,height=7cm,angle=-90}}
\vspace{0.5cm}
{\footnotesize Fig. 1.
Phase space diagram of the damped pulsating oscillator. \\
We choose the parameters to be $m_0=1=\Omega$, $\gamma=1/10$, $\mu=4$,
and $\nu=1/3$. \\
(a) The damped-forced oscillator with the external force $F(t)=\sin(t)$.\\
(b) The damped oscillator without external force.\\
Each initial state is the coherent state with $\left<q(0)\right>=5, 
\left<p(0)\right>=0$.
Each real line denotes the time evolution of $\left<q(t)\right>$ and 
$\left<p(t)\right>$.
The ellipses are depicted for every four timing units to denote 
$\left<\Delta q(t)\right>$~(horizontal axis) and 
$\left<\Delta p(t)\right>$~(vertical axis).
Note that two ellipses of (a) and (b), with the same time, have the
same shapes.
}
\end{figure}

\section{Summary}

We considered the time-dependent harmonic oscillator
with a driving force and found its quantum solutions.
We found the LR invariant and
constructed the Fock space as its eigenstates.
The quantum states of Lo~\cite{lo} can be constructed by
squeezing and displacing our quantum states
as seen from (\ref{0B:0a}).
Further, our ground state is the coherent state
of the invariant with no external force -- see (\ref{0B:0b}).
We showed that the ground state of the invariant has
lower energy expectation value than the ground state
of the instantaneous Hamiltonian.

Using the time evolution of the creation and annihilation operators
for the LR invariant, we also found the solutions of the Heisenberg
equation of motion for the position and momentum operator.
The exact quantum motion of a {\it forced} time-dependent
harmonic oscillator is described by the classical solution of
the corresponding {\it unforced} oscillator.
The dispersions of $q$ and $p$ for the eigenstates and
the coherent state of the invariant
do not depend on the external force.
It was found that the external force merely shifts the
wavepackets of the eigenstates of the invariant
{\it with no external force}.
These results were exemplified by a model.

\section*{acknowledgments}
This work was supported in part by the
Korea Science and Engineering Foundation (KOSEF).


\begin{references}
\bibitem{jann}
	A. D. Jannussis, G. N. Brodimas, and A. Streclas,
	Phys. Lett. A {\bf 74}, 6(1979).
\bibitem{lewis}
	H. R. Lewis. Jr., and W. B. Riesenfeld, J. of Math. Phys. {\bf 10},
	1458(1969).
\bibitem{sangpyo}
	S. P. Kim, J. Phys. A Math. Gen. {\bf 27}, 3927(1994).
\bibitem{jyji1}
	J. Y. Ji, J. K. Kim, and S. P. Kim, Phys. Rev. A
	{\bf 51}, 4268(1995).
\bibitem{kha}
	D. C. Khandekar and S. V. Lawande,
	J. Math. Phys. {\bf 20}, 1870(1979).
\bibitem{lo}
	C. F. Lo, Phys. Rev. A {\bf  43}, 404(1991); Phys. Rev. A {\bf 45}
	5267(1992).
\bibitem{jyji3}
	J. Y. Ji, and J. K. Kim, Phys. Rev. A {\bf 53},
	(to appear) (1996).
\end{references}
\end{document}